\documentclass[useAMS,usenatbib]{mn2e}

\def\Msol{\hbox{M$_\odot$}}
\def\Msun{\hbox{M$_\odot$}}

\def\kms{\hbox{km$\,$s$^{-1}$}}
\def\cmt{\hbox{cm$^{-3}$}}

\def\two{\,{\sc ii}}
\def\three{\,{\sc iii}}

\def\fsec{\hbox{$.\!\!^{\rm s}$}}

\usepackage{amsmath}
\usepackage{amssymb}
\usepackage{mathrsfs}
\usepackage{color}
\usepackage{upgreek}
\usepackage{graphicx}

\title[Wind--clump interactions in NGC 6357]{VLT/FLAMES-ARGUS observations of stellar wind--ISM cloud interactions in NGC 6357\thanks{Based on observations collected at the European Organisation for Astronomical Research in the Southern Hemisphere, Chile under programme 081.C-0808.}}
\author[M.S.\ Westmoquette et al.] {M.\ S.\ Westmoquette$^1$\thanks{E-mail: msw@star.ucl.ac.uk}, J.\ D.\ Slavin$^2$, L.\ J.\ Smith$^{1,3,4}$, J.\ S.\ Gallagher III$^{5}$ \\
$^1$Department of Physics and Astronomy, University College London, Gower Street, London, WC1E 6BT, UK\\
$^2$Harvard-Smithsonian Center for Astrophysics, 60 Garden St., MS 83, Cambridge, MA 02138, USA\\
$^3$Space Telescope Science Institute, 3700 San Martin Drive, Baltimore, MD 21218, USA\\
$^4$European Space Agency, Research and Scientific Support Department, Baltimore, MD 21218, USA\\
$^5$Department of Astronomy, University of Wisconsin-Madison, 5534 Sterling, 475 North Charter St., Madison, WI 53706, USA\\
}
\date{Accepted 2009 October 20. Received 2009 October 19; in original form 2009 September 14}
\pagerange{\pageref{firstpage}--\pageref{lastpage}}
\pubyear{2009}
\begin{document}
\maketitle
\label{firstpage}

\begin{abstract}
We present optical/near-IR integral field unit (IFU) observations of a gas pillar in the Galactic H\two\ region NGC 6357 containing the young open star cluster Pismis 24. These observations have allowed us to examined in detail the gas conditions of the strong wind-clump interactions taking place on its surface.

By accurately decomposing the H$\alpha$ line profile, we identify the presence of a narrow ($\sim$20~\kms) and broad (50--150~\kms) component, both of which we can associate with the pillar and its surroundings. Furthermore, the broadest broad component widths are found in a region that follows the shape of the eastern pillar edge. These connections have allowed us to firmly associate the broad component with emission from ionized gas within turbulent mixing layers on the pillar's surface set up by the shear flows of the winds from the O stars in the cluster. We discuss the implications of our findings in terms of the broad emission line component that is increasingly found in extragalactic starburst environments. Although the broad line widths found here are narrower, we conclude that the mechanisms producing both must be the same. The difference in line widths may result from the lower total mechanical wind energy produced by the O stars in Pismis 24 compared to that from a typical young massive star cluster found in a starburst galaxy.

The pillar's edge is also clearly defined by dense ($\lesssim$5000~\cmt), hot ($\gtrsim$20\,000~K), and excited (via the [N\two]/H$\alpha$ and [S\two]/H$\alpha$ ratios) gas conditions, implying the presence of a D-type ionization front propagating into the pillar surface. Although there must be both photoevaporation outflows produced by the ionization front, and mass-loss through mechanical ablation, we see no evidence for any significant bulk gas motions on or around the pillar. We postulate that the evaporated/ablated gas must be rapidly heated before being entrained.
\end{abstract}



\section{Introduction} \label{sect:intro}

\subsection{Broad emission features}
Direct evidence of interactions between stellar winds and their surroundings is relatively sparse. Observations of H$_{2}$ and CO absorption lines on sight-lines towards massive stars in our Galaxy show comparatively broad underlying components \citep*{falgarone90, hartquist92}. These were interpreted as arising in turbulent surface layers on cold dense molecular clouds surrounding the stars resulting from the impact of the stellar winds. Broad underlying components have also been observed in the optical emission lines from more energetic environments in nearby giant H\two\ regions \citep[e.g.][]{roy92, chuken94, yang96} and starburst galaxies (e.g.~\citealt{homeier99, marlowe95, mendez97}; \citealt*{sidoli06}; \citealt{vanzi06}). However, due to mismatches in spectral and spatial resolution and in the specific environments observed, the nature of the energy source for these broad optical lines has been contested. Recent detailed integral field unit (IFU) studies of the ionized interstellar medium (ISM) in the nearby starburst galaxies NGC 1569 and M82, however, have shed a considerable amount of light on this problem \citep{westm07a, westm07b, westm08, westm09a, westm09b}.

By mapping out the properties of the individual emission line components in NGC 1569 (including a broad underlying component with FWHM $\sim$ 100--350~\kms), \citet{westm07a} identified a number of correlations that allowed them to determine the likely origin of the broad component. They concluded that, like in the CO line case, the evaporation and/or ablation of material from cool interstellar gas clouds caused by the impact of the high-energy photons and fast-flowing cluster winds \citep{pittard05} produces turbulent mixing layers (TMLs) on the surfaces of the clouds \citep{begelman90, slavin93, binette99, binette09} from which the emission arises. \citet{westm07c, westm09a, westm09b} showed that this explanation is also applicable to the broad emission lines detected in M82. They argued that since M82's high pressure ISM is highly fragmented with many small clouds well mixed in with the star clusters, there are many cloud surfaces with which the copious ionizing photons and fast winds can interact and this can explain the prominent broad line emission.

\citet{binette09} modelled the H$\alpha$ and [O\three]$\lambda$5007 broad emission components observed in the extragalactic giant H\two\ region NGC 2363 with a TML model. By using an approximate treatment of turbulence \citep[the mixing length approach of][]{canto91}, they avoided dealing with the details of the turbulent energy cascade, while still being able to predict the line kinematics. The \textsc{Mappings Ic} code \citep{ferruit97} was used to compute the radiative transfer and thus the line emissivities. In this model the line width is created by the velocity gradient in the mixing layer. A limitation of this approach is that the turbulent velocity that should be generated in the TML is not included and neither are the non-equilibrium ionization (NEI) effects that are expected when cooler gas mixes rapidly with hotter gas. This model is therefore only a good approximation if the mixing occurs fairly slowly and the ionization is dominated by photoionization. The model of \citet{slavin93} includes NEI effects, but does not predict line profiles since the dynamics in the layer are not modelled. 

The only way to connect the low-energy molecular line cases to the higher-energy optical emission line cases is to model the physical process using the correct initial conditions. We have begun to do this using a hydrodynamical code that includes NEI effects, but we must first confirm that the physical mechanism producing the broad component in both cases is wind-driven TMLs, and we must have reliable measurements of the gas conditions within these layers. Thus we have undertaken a study of a nearby Galactic analogue to the type of environments found in optical studies of extragalactic giant H\two\ regions and starburst galaxies. Although the energies and ISM conditions found in starburst galaxies are not easily found within our Galaxy, the advantage of proximity allows discrete superimposed kinematical components to be clearly resolved and unambiguously disentangled from other possible line broadening mechanisms.

\begin{figure*}
\centering
\includegraphics[width=\textwidth]{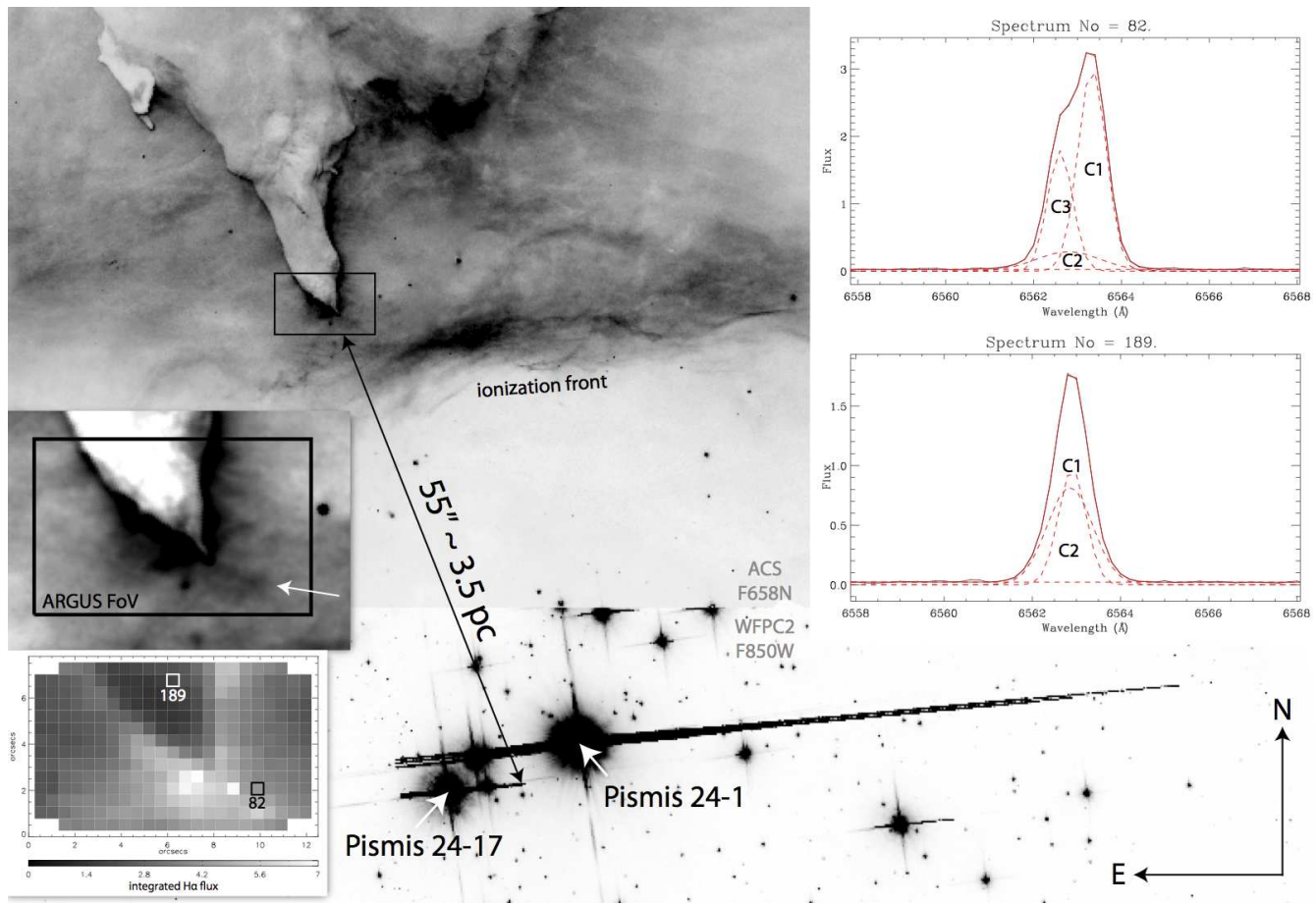}
\caption{Composite image of NGC 6357: the upper half comprises an ACS+WFPC2 F658N (H$\alpha$) composite. Since this image does not cover the stars of the cluster, we have extended the figure southwards by including part of the \textit{HST}/ACS WFC F850W ($\sim$$I$-band) image. The ARGUS IFU position is outlined with a solid rectangle ($11.5\times 7.3$ arcsec), the two main ionizing stars \citep{walborn02, bohigas04} and the ionization front are labelled, and the approximate projected distance from the stars to the tip of the pillar is indicated. \textit{Inset left, top:} zoom-in showing the outline of the IFU position on the pillar. The image intensity scale has been stretched to enhance the protrusion indicated by the arrow and discussed in the text. \textit{Inset left, bottom:} integrated H$\alpha$ flux (C1+C2+C3) map as seen through the IFU. H$\alpha$ line profiles and Gaussian fits from the two spaxels indicated are shown in the \textit{top right inset}, where the flux units are in an arbitrary but relative scale.}
\label{fig:finder}
\end{figure*}

\subsection{NGC 6357}

The analogue we have chosen is a gas pillar that forms part of the youngest H\two\ region (G353.2+0.9) in NGC 6357 (RCW 131, W 22, Sh-2 11), a large H\two\ region/molecular cloud complex in the Sagittarius spiral arm of the Milky Way. G353.2+0.9 is ionized by the open cluster Pismis 24. This cluster has a stellar population a few times that of the Orion star cluster and a central stellar density similar to that of the Galactic young massive cluster NGC 3603 \citep{wang07}. Pismis 24 is still too young to have had any supernova explosions \citep{wang07}; this is in-keeping with its O-star population, which includes an O3.5 If (Pismis 24-1) and O3.5 III(f*) (Pismis 24-17) star \citep{walborn02}, each of $>$100~\Msol\ \citep{maiz07}.

The conclusion of a number of individual radio and IR studies of NGC 6357 is that most of the molecular material is located either behind the H\two\ region and/or north of the sharp ionization front clearly visible in Fig.~\ref{fig:finder} \citep{massi97}. The expansion of the H\two\ region has caused gas to accumulate in a (large) interim zone into which the ionization front is proceeding; this has triggered the formation of several more (less massive) stars within the cloud \citep{bohigas04}. The molecular gas observations suggest that NGC 6357 represents a late-stage blister-type H\two\ complex viewed face-on, meaning that we are seeing the top ionized face of a much larger molecular cloud \citep{massi97}, and that the pillar we have focussed on must project up from this structure towards the observer. Since the most highly excited regions of the nebula are facing the cluster, it is thought that the primary ionizing sources must be the two O3.5 stars in the Pismis 24 cluster. This is confirmed by ionization calculations that show that the exciting stars must be of spectral type earlier than O5 \citep{bohigas04}. The location of these stars and their projected distance to the gas pillar and ionization front (3.5~pc) are indicated in Fig.~\ref{fig:finder}. 

The mechanisms that lead to the formation of ``elephant trunks'' in H\two\ regions is still a matter of debate \citep[e.g.][]{mizuta06, gritschneder09}. However, the fact that these pillar-shaped nebulae usually point towards and exhibit velocity gradients in the direction of a nearby OB star, and show signs of compression and/or star formation at their tips, suggests a strong interaction with the winds and radiation from these sources \citep{pound98, sugitani02, gahm06, mizuta06, gritschneder09}. Our targeted pillar in NGC 6357 both points directly towards (in projection) the aforementioned O stars in Pismis 24 (Fig.~\ref{fig:finder}), and has been detected in both CO and HCO$^{+}$ implying that it contains very high density gas \citep[$>$$10^{5}$--$10^6$~\cmt;][]{massi97}. The pillar is also known to house two compact radio sources thought to be young stellar objects, including one at its apex \citep[IRS 4;][]{persi86, felli90, wang07}.


In this paper we present optical/near-IR VLT/FLAMES-ARGUS IFU observations of the tip of this gas pillar (Section~\ref{sect:obs}). In these observations we have identified multiple broad and narrow emission line components. This has allowed us to examine in detail the kinematics (radial velocity and FWHM) of each H$\alpha$ line component, and the spatially resolved nebular properties (densities, temperatures, excitations) of this region (Section~\ref{sect:maps}), and use these insights to discuss the wind and ionization interactions taking place on the pillar's surface (Section~\ref{sect:disc}). A summary of these findings and discussion is contained in Section~\ref{sect:conc}.
We adopt a distance to the NGC 6357 complex of 2.56~kpc \citep{massey01}, meaning 10 arcsec $\sim$ 0.62~pc.

\section{Observations and data reduction} \label{sect:obs}
Observations of NGC 6357 were obtained with the FLAMES instrument on the VLT with the ARGUS IFU coupled to the GIRAFFE spectrograph (PI: Westmoquette, Prop ID: 081.C-0808). We used the ARGUS array in its 0.52 arcsec/spaxel spatial sampling mode, giving a field-of-view (FoV) of $11.5\times 7.3$ arcsec sampled by 333 fibres, and centred it on $17^{\rm h}\,24^{\rm m}\,45\fsec8$, $-34^{\circ}\,11'\,04\farcs1$ (J2000) in order to cover the tip of the pillar. Image quality during the observations fell in the range 0.7--1.0$''$. We observed this position with three grating settings (L614, L682 and L881; giving wavelength coverages of 5750--7200\,\AA\ and 8200--9400\,\AA) with exposure times of 2100--2700~secs. For each grating setting, two individual exposures were taken with a dither offset of $0\farcs52$ in RA (1 fibre) in order to facilitate the masking of the dead fibres. Fig.~\ref{fig:finder} shows the position of the IFU on an \textit{HST} ACS and WFPC2 F850W ($\sim$$I$-band) and F658N (H$\alpha$) composite image of NGC 6357, together with the location and distance of the primary ionizing sources of this nebula \citep[Pismis 24-1 and 24-17;][]{bohigas04}.

Basic data reduction was achieved with the girBLDRS (Giraffe BaseLine Data Reduction Software). This included bias subtraction, flat fielding, identification and spectrum extraction, and wavelength calibration. Post-processing was carried out in {\sc iraf}\footnote{The Image Reduction and Analysis Facility ({\sc iraf}) is distributed by the National Optical Astronomy Observatories which is operated by the Association of Universities for Research in Astronomy, Inc. under cooperative agreement with the National Science Foundation.}, including cosmic-ray rejection \citep[using \textsc{lacosmic};][]{vandokkum01}, flux calibration and the combining of the individual exposures (dead fibres were masked and removed when the dithered exposures were combined).

The FWHM spectral resolution, measured from averaging fits to isolated arc lines over all spaxels, was found to be 0.55~\AA\ (L614 grating), 0.55~\AA\ (L682 grating), and 0.89~\AA\ (L881 grating).

\subsection{Emission line fitting} \label{sect:fitting}
The high S/N and spectral resolution of the data have allowed us to quantify the emission line profile shapes to a high degree of accuracy. In general, we find the emission lines to be composed of two bright, narrow components (FWHM $\sim$ 10--30~\kms; hereafter C1 and C3) overlaid on a fainter, broad component (FWHM $\sim$ 40--150~\kms; hereafter C2). Following the methodology first employed by \citet{westm07a}, we fitted multiple Gaussian profile models to each emission line using an \textsc{idl}-based $\chi^{2}$ fitting package called \textsc{pan} \citep[Peak ANalysis;][]{dimeo}. Each line in each of the 333 spaxels was fitted using a single-, double- and triple-Gaussian initial guess. Line fluxes were constrained to be positive and widths to be greater than the instrumental contribution to guard against spurious results. To fit the [S\two] doublet, for each component we constrained the wavelength difference between the two Gaussian models to be equal to the laboratory difference, and FWHMs to equal one another. Multi-component fits were run several times with different initial guess configurations (widths and wavelengths) in order to account for the varied profile shapes, and the result with the lowest $\chi^2$ was retained.

Determining, firstly, how many Gaussian components best fit an observed profile, and secondly, which fits are most appropriate, is not trivial. To decide how many Gaussian components best fit an observed profile (one, two or three), we employed the statistical F-test. This ascertains if the $\chi^2$ improvement obtained when increasing the number of fit components is statistically significant or not \citep[see][for a fuller description of this procedure]{westm07a}.

Experience has taught us that a number of additional, physically motivated tests are needed to firstly filter out well-fit but physically improbable results, and secondly to assess which Gaussian profile belongs to which physical line component. For a fit to be accepted, we set the criteria that the measured FWHM had to be greater than the associated error on the FWHM result (a common symptom of a bad fit). Of a triple-Gaussian fit, we specified that the broadest component should be assigned to component 2 (C2), and after that, the redder to be component 1 (C1) and the bluer to be component 3 (C3). The reasons for this choice will become apparent in the following sections. With double-Gaussian fits, we assigned C2 to be the broader of the two components and C1 to the other, regardless of their velocity difference. This consistent approach helped limit the confusion that might arise during analysis of the results where discontinuous spatial regions might arise from incorrect component assignments. The top-right inset to Fig.~\ref{fig:finder} shows an example of a triple- and double-Gaussian H$\alpha$ line profile extracted from the data cube, and illustrates how the components were assigned.


\begin{figure}
\centering
\includegraphics[width=0.3\textwidth]{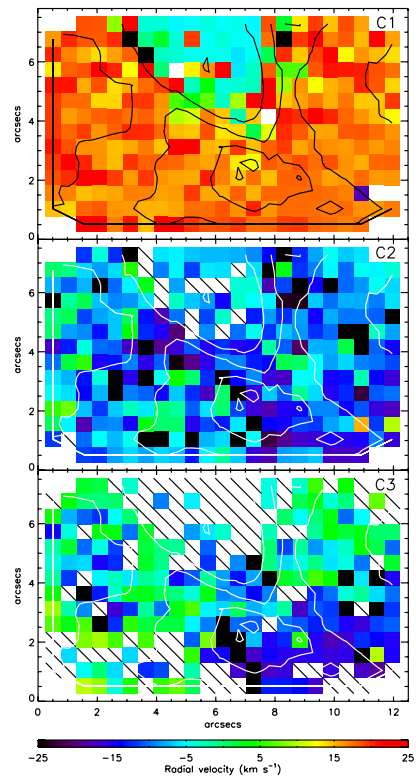}
\caption{H$\alpha$ radial velocities in units of \kms\ (corrected to $v_{\rm{LSR}}$) for the three line components. Contours represent the summed (C1+C2+C3) H$\alpha$ flux as mapped in the bottom-left inset to Fig.~\ref{fig:finder}.}
\label{fig:Ha_vel}
\end{figure}

\begin{figure}
\centering
\includegraphics[width=0.3\textwidth]{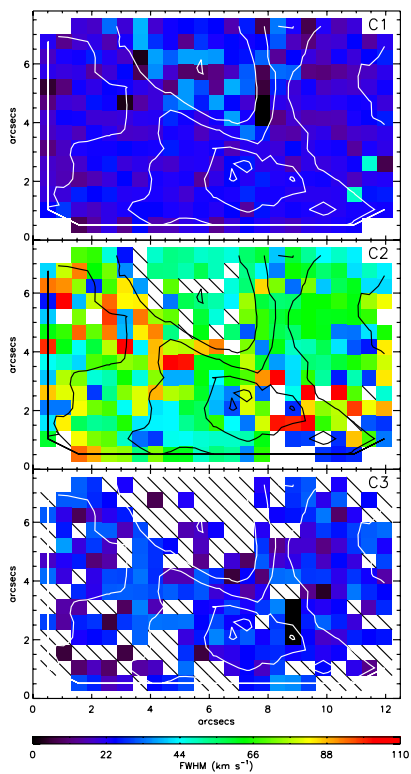}
\caption{H$\alpha$ FWHM in units of \kms\ (corrected for instrumental broadening) for the three line components, C1, C2 and C3. Contours represent the summed H$\alpha$ flux shown in the bottom-left inset to Fig.~\ref{fig:finder}.}
\label{fig:Ha_fw}
\end{figure}

\begin{figure*}
\centering
\includegraphics[width=0.9\textwidth]{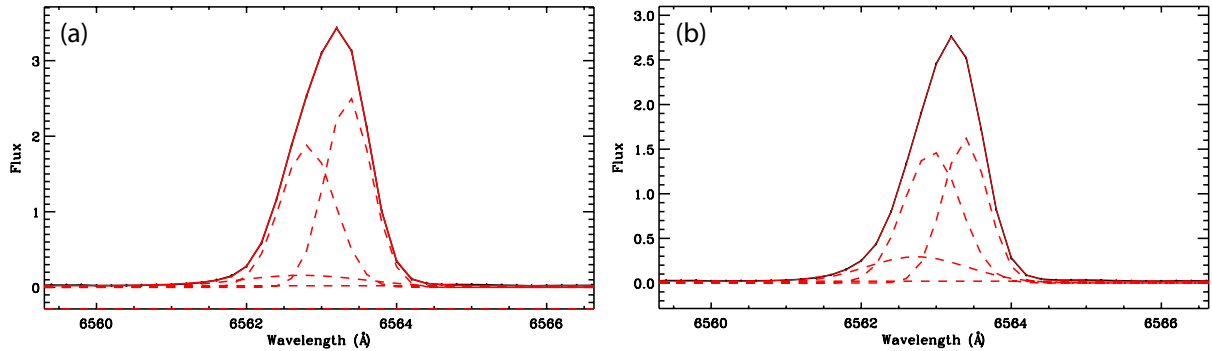}
\caption{Average H$\alpha$ line profiles and corresponding Gaussian fits from (a) the diagonal N-E to S-W strip containing the broadest broad line emission (total 55 spaxels) -- C2 (broad component) FWHM = 87~\kms; (b) the remaining area (total 278 spaxels) -- C2 FWHM = 68~\kms. 
}
\label{fig:average_profiles}
\end{figure*}

\section{Emission line maps} \label{sect:maps}

\subsection{H$\alpha$ kinematics}
Maps of the H$\alpha$ radial velocity (corrected to $v_{\rm LSR, kinematic}$) and FWHM (corrected for instrumental broadening) for the three identified line components are shown in Figs.~\ref{fig:Ha_vel} and ~\ref{fig:Ha_fw}. The radial velocities are shown relative to $v_{\rm LSR}$ to aid comparison with the CO measurements of \citet{massi97}. On each map, contours of the summed (C1+C2+C3) H$\alpha$ flux distribution are shown to locate the tip of the pillar and help guide the eye.

These maps reveal a number of important characteristics about the ionized gas dynamics. South of the pillar the H$\alpha$ line is split into a bright, redshifted, narrow component (C1), a fainter, narrow component (C3) with velocities ranging from +5 to $-20$~\kms, and a broad, mostly blueshifted component (C2). The top panel of the right-hand inset to Fig.~\ref{fig:finder} shows an example line profile from this region (Spectrum 82).

H$\alpha$ C3 exhibits a velocity gradient from the south-west to the north and east of the field, ranging from $-20$ to +5~\kms. The CO velocity of gas surrounding the pillar \citep[$\sim$0~\kms;][]{massi97} is in agreement with the reddest of these velocities, suggesting that there is some correspondence between the molecular and ionized gas, although the bluest C3- and the majority of the C2-emitting ionized gas must certainly be flowing towards the observer at velocities up to $\sim$20~\kms\ relative to the CO gas. No evidence of the redshifted C1 component can be seen in the CO velocity maps of \citet{massi97}. We caution, however, that the spatial resolution of the CO observations is $\sim$40$''$, meaning that a detailed comparison is meaningless. 

Moving from here onto the pillar surface, the bright redshifted component disappears and the remaining narrow component now becomes assigned to C1 (hence the apparent discontinuity in the C1 map). The bottom panel of the right-hand inset to Fig.~\ref{fig:finder} shows an example line profile from this region (Spectrum 189). Here the velocity of C1 is consistent with the velocity of C3 in spaxels adjacent to this region, leading us to conclude that the component that was assigned to C3 in the spaxels surrounding the pillar is in fact the same component as that which becomes re-assigned to C1 on the pillar (as the redshifted component is now obscured). The velocity of both the narrow (C1) and broad (C2) H$\alpha$ components on the pillar are in very good agreement with that of the corresponding CO velocities \citep[$v_{\rm LSR} \sim -5.5$~\kms;][]{massi97}, indicating that the ionized and molecular gas in the pillar must be physically contiguous. The similarity of the broad component velocities with the bluer narrow component (C3 surrounding the pillar and C1 on the pillar) suggests that C2 is indeed emitted by turbulent gas associated with the pillar. We discuss this further in Section~\ref{sect:disc}.

From Fig.~\ref{fig:Ha_fw}, the width of the narrow components
(C1 and C3) appears very constant over the whole field-of-view (FoV). The mean FWHM of the two narrow components is $20\pm 7$~\kms\ (where the quoted uncertainty represents the standard deviation on the sample and quantifies the degree of consistency). The broad component (C2; FWHM $>$ 40~\kms) is identified in the majority of spaxels across the FoV, and rises to widths of $\sim$150~\kms\ in places\footnote{The broad component can also identified in the [S\three]$\lambda$9069 line but at a lower signal-to-noise (S/N).}. The broadest H$\alpha$ C2 lines (FWHM $\sim$ 90--150~\kms) are found along a diagonal strip extending from the north-east of the FoV, tracing the eastern edge of the pillar structure, along to the tip, then extending off towards the south-west. This coincidence lends further support to the association of the C2-emitting gas with the pillar surface. Close inspection of the \textit{HST} H$\alpha$ image reveals a very low surface brightness protrusion to the pillar tip that is coincident with the south-west extension to the diagonal strip of broad line emission. This is highlighted in the left-hand inset to Fig~\ref{fig:finder} and discussed further in Section~\ref{sect:disc}. Fig.~\ref{fig:average_profiles} shows the average H$\alpha$ line profile and the associated multi-component Gaussian fits for the 55 spaxels in the diagonal N-E to S-W strip containing the broadest broad line emission (panel a) and the remaining 278 spaxels (panel b). Although on average the broad component in this diagonal strip is fainter than in the rest of the field, it is clearly significantly broader (87~\kms\ vs.\ 68~\kms).

\subsection{Nebular properties}
The integrated (C1+C2+C3) H$\alpha$ line flux map is shown in the bottom-left inset to Fig.~\ref{fig:finder}. The vast majority of the H$\alpha$ flux across the IFU field originates from C1 and C3 (and then mostly C1); C2 makes a very small contribution to the total line flux [$\lesssim$ 0.3 $\times$ (C1+C3)]. On the pillar, however, the narrow line (C1) is much fainter, and C2 becomes more dominant, but in some places C2 and C1 make an almost equal contribution to the total line flux.

\begin{figure*}
\centering
\includegraphics[width=0.7\textwidth]{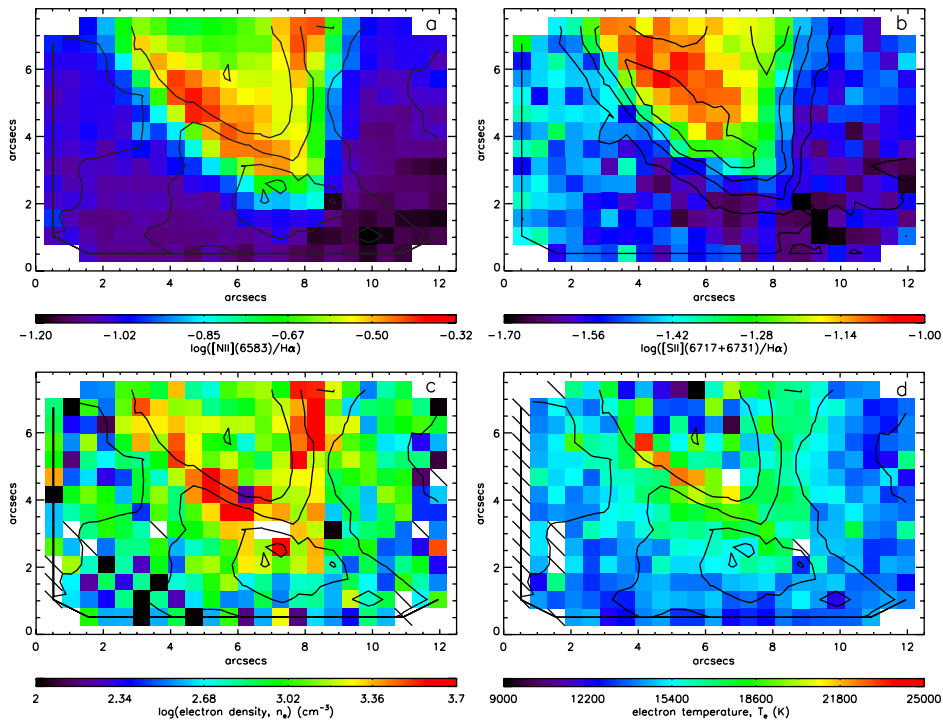}
\caption{(a) log([N\two]$\lambda$6583/H$\alpha$) ratio; contours represent the corresponding summed H$\alpha$ fluxes (bottom-left inset to Fig.~\ref{fig:finder}). (b) log([S\two]$\lambda$6717+$\lambda$6731)/H$\alpha$ ratio. The contours represent the [N\two]/H$\alpha$ map and are shown to highlight spatial difference between the peak [S\two]/H$\alpha$ ratios and the peak [N\two]/H$\alpha$ ratios. (c) electron density map; contours represent the summed H$\alpha$ fluxes. (d) electron temperature map derived from the dereddened [S\three]($\lambda$9069+$\lambda$9532)/$\lambda$6312 flux ratio. Each map has been calculated using summed fluxes from all identified line components (i.e.\ C1+C2+C3).}
\label{fig:ratio_elec_comb}
\end{figure*}

Maps of the [N\two]$\lambda$6583/H$\alpha$ and [S\two]$\lambda$6717+$\lambda$6731)/H$\alpha$ line ratios are shown in Figs.~\ref{fig:ratio_elec_comb}a and \ref{fig:ratio_elec_comb}b. Although neither indicators show any evidence for shock-like ratios [log([N\two]/H$\alpha$)$\gtrsim$$-0.2$ or log([S\two]/H$\alpha$)$\gtrsim$$-0.4$], both ratios are significantly enhanced on the pillar compared to the surrounding gas. In [N\two]/H$\alpha$ (Fig.~\ref{fig:ratio_elec_comb}a), the peaks are located on the south-western and eastern pillar edges, whereas in [S\two]/H$\alpha$ (Fig.~\ref{fig:ratio_elec_comb}b) the peak is located nearer the pillar's central axis, interior to the [N\two]/H$\alpha$ peaks. This difference is made clear by comparing the [S\two]/H$\alpha$ map to the overplotted contours representing the [N\two]/H$\alpha$ of Fig.~\ref{fig:ratio_elec_comb}a. Since the [S\two] line flux may be enhanced by emission from partially ionized regions, we might expect [S\two] to originate from deeper within the cloud compared to [N\two]. The line ratio maps are certainly consistent with this idea. We might then also expect [S\two] to exhibit a narrow line width, since the interior regions of the clump are presumably less turbulent. To investigate this we summed 20 spaxels covering the peak in [N\two]/H$\alpha$ and 20 spaxels covering the [S\two]/H$\alpha$ peak and measured the widths of the [N\two] and [S\two] lines in the two resulting spectra. To within the uncertainties, we find the line widths are equal and do not exhibit any significant change between the two locations. Observations with a S/N ratio sufficient to accurately decompose the [N\two] and [S\two] lines would be needed to investigate this in more detail.

A map of the electron densities, calculated from the flux ratio of the [S\two]$\lambda \lambda$6717,6731 doublet (assuming $T_{\rm e}=10^4$~K), is shown in Fig.~\ref{fig:ratio_elec_comb}c. Here we sum the fluxes from all components in the [S\two] lines to increase the S/N ratio. The density of the gas projected on the body of the pillar is $\sim$1000--1500~\cmt, only marginally higher than that of the gas surrounding the pillar (500--1000~\cmt). These values are not unsurprisingly high for a large, gas-rich H\two\ region. However, the measured densities at the pillar edges are significantly enhanced, peaking at $\sim$5000~\cmt.

Electron temperatures were derived from the dereddened flux ratio of [S\three]($\lambda$9069+$\lambda$9532)/$\lambda$6312. To deredden this ratio, we first calculated the extinction in each spaxel using an average of the P11/H$\alpha$, P10/H$\alpha$, P9/H$\alpha$ decrements, using the total summed flux (C1+C2+C3) from each line, the \citet{cardelli89} extinction curve assuming $R_{V}$=3.5 \citep{bohigas04}, and the electron densities calculated above. We find extinctions in the range $A_{V}$=3--5 mags, which is consistent with previous studies \citep[e.g.][]{bohigas04}. Since our observations only cover [S\three]$\lambda$6312 and [S\three]$\lambda$9069, we assumed [S\three]$\lambda$9532/$\lambda$9069 = 2.48 \citep{mendoza82} to recover the [S\three]$\lambda$9532 fluxes. The resulting electron temperature map is shown in Fig.~\ref{fig:ratio_elec_comb}d. Electron temperatures of $\sim$12\,000--14\,000~K are found for the gas surrounding the pillar. The temperature then clearly increases towards the edges of pillar, peaking on the eastern edge at values of $\gtrsim$20\,000~K. The temperature of the gas projected on the centre of the pillar ($\sim$9000--10\,000~K) is lower than the surrounding gas.

In summary, our excitation ([S\two]/H$\alpha$ and [N\two/H$\alpha$), electron density ([S\two] doublet), and electron temperature ([S\three][$\lambda$9069+$\lambda$9532]/$\lambda$6312) diagnostics reveal that the gas on the pillar edge is clearly more highly excited, denser ($\sim$5000~\cmt\ compared to $\sim$1000~\cmt), and hotter (20\,000~K \textit{vs.} 12\,000--14\,000~K) than its surroundings. This suggests that the gas conditions change dramatically within the surface layers of the pillar.

\section{Discussion} \label{sect:disc}

The dense, hot, and highly excited conditions of the gas on the pillar's edge are consistent with what might be expected after a D-type ionization front has propagated into the pillar \citep{osterbrock06}. A D-type ionization front can occur when a supersonic ionization front runs into a dense neutral gas cloud. The increase in density causes the ionization front to slow down enough to allow the ionized gas behind the front time to move in response to the heating. The gas velocity quickly becomes supersonic, and the ram pressure of this material streaming away from the front towards the star (down the density gradient) causes a shock wave to break off from the ionization front and begin to compress the gas ahead of the front. The shock front gradually weakens and the ionization front continues into the clump as a strong D-type front with a large density jump.

The density and temperature increases we observe on the pillar surface are consistent with a radiative shock, as would be expected in such a dense environment. In this case the degree to which these properties increase is governed by the magnetic field strength that acts to limit compression and dominate the post-shock pressure. The material streaming away from the cloud, either as a result of the aforementioned ram pressure or from the ionization and accompanying heating, forms what is known as a photoevaporation flow.

Besides photoevaporation and TMLs, hydrodynamic models of wind--clump interactions also predict that material should be stripped (ablated) from the surface layers by the impact of the stellar wind and incorporated (entrained) into the wind flow \citep{hartquist86, marcolini05}. However, disregarding the redshifted narrow component presumed to originate from background gas, we see no evidence for any significant bulk gas motions on or around the pillar (such as split line profiles or diverging velocity gradients). The radial velocity of the narrow and broad components are furthermore consistent with that of the CO gas indicating that the ionized and molecular gas are physically contiguous. This lack of bulk motions could be for a number of reasons: no evaporation/ablation and entrainment of material is taking place; the bulk gas motions are parallel to the pillar surface and therefore in the plane of the sky and not visible in the radial velocity maps; or the gas is rapidly heated to $T\gg 10^4$~K before being entrained into the flow making it invisible in the optical. That no outflow and entrainment of material is taking place is highly unlikely given the clear evidence of a strong ionization front and the mechanical energy of the winds from the O stars that must be impinging on the clump surface. It is also unlikely that the gas motions are all sufficiently parallel to the pillar surface not to be visible, so we have to conclude that the evaporated/ablated gas must be rapidly heated before being entrained (i.e.\ the gas heating time is fast compared to outflow time). Thermal conduction-driven heating could help here since this process is known to be able to rapidly heat material so that it quickly becomes part of the ambient flow \citep{pittard07}.

Could the large turbulent velocities we observe on the outer, optically-emitting surfaces of the pillar drive turbulence deeper into the pillar core? In their molecular line observations of the pillar, \citet{massi97} report CO line widths of 2--3~\kms. These are indeed highly suprathermal for a molecular gas at their measured temperature of $\sim$30~K (where the sound speed would be $\sim$0.5~\kms). Furthermore, the reported line widths of HCO$^{+}$, a species that traces higher density gas, are much narrower, suggesting that the the degree of turbulence must fall off rapidly with depth into the pillar. While it is conceivable that the large turbulent velocities that we observe on the outer surfaces of the pillar drive the turbulence of these deeper, denser layers, the large difference in velocities (100~\kms\ \textit{vs.} 3~\kms) implies that the situation may not be as straightforward as that. We intend to examine this question in the future with detailed hydrodynamical models.

Contrary to expectation the region of broadest H$\alpha$ C2 line widths does not wrap around the western edge of the pillar, but continues from the pillar tip in the south-west direction, following the faint protrusion to the pillar (highlighted in Fig.~\ref{fig:finder}). If this region of broadest C2 emission traces the TML set up by the shear stellar wind flow, then the strongest wind--clump interaction must be taking place along this strip. It is interesting that the protrusion does not stand out in any of our nebular diagnostic maps; we attribute this to the fact that emission from the protrusion is too faint to be detected in any line but H$\alpha$ in our spectra. Deeper observations would be needed to confirm this.

What might this protrusion be? Two possibilities are: (1) it is a part of the original molecular gas cloud from which Pismis 24 formed that remains to be eroded or (2) that it has formed as a result of the hydrodynamics of the wind-clump interaction. If there were some outflow from the pillar as a result of ionization/heating, then in general it would flow away from the high pressure zone where the heating is occurring, along the sides of the clump and be swept downstream by the wind. A highly unstable situation would occur at the tip, where the pressure would push the outflow directly into the wind, and, for a time before the ram pressure of the wind overwhelms that of the outflow, could conceivably produce a protrusion such as we see. The geometry might be such that this protrusion is shielding the western edge of the main pillar body from the oncoming winds therefore preventing the development of a TML on its surface. 

An alternative idea is that the protrusion represents one half of a jet \citep[or a monopolar jet;][]{henney02} from the IR and X-ray detected YSO known to inhabit the apex of the pillar, IRS 4 \citep[likely a B0--B2 protostar;][]{persi86, wang07}; in this case the turbulent line widths would arise from interactions between the jet and the surrounding gas. Although the length of the protrusion ($\sim$2.6$''$ $\approx$ 3300~AU) is plausible for a microjet \citep{bally00}, the fact that the structure appears curved and that we see no evidence for it in the radial velocity maps makes this argument very unlikely. We therefore favour the former scenario where the structure is formed from a remnant overdensity in the ambient medium or an instability in the outflowing material.

%
The existence of a broad component, not just on the pillar surface, but across the entire IFU field-of-view indicates that turbulent $10^4$~K gas pervades the entire region. Although detailed hydrodynamic models would be needed to ascertain whether this gas exists as a result of the evaporation/ablation flows from the pillar surface, the significant difference in width between the broad component associated with the TML ($>$100~\kms) and the surrounding gas ($\sim$50~\kms), together with the excitation and density results, suggests that these two gas bodies are not related.
Can these results help us interpret what we see in more distant and energetic starburst systems \citep[e.g.][]{westm07a, westm07b, westm09a, westm09b}? The C2/C1 flux ratio is, in general, similar to what we have found in starburst galaxies, suggesting that the TMLs in both cases are of similar importance. The C2 widths, however, are narrower (50--150~\kms\ \textit{vs.} 150--350~\kms). The degree of turbulence in the TML might be expected to depend on the ratio of mass flux in the wind to that in the ionization initiated outflow from the clump. We know the Pismis 24 cluster contains only a few O stars [Pismis 24-1 -- which \citet{maiz07} resolve into three objects with a combined mass of $\sim$200~\Msun\ -- and Pismis 24-17] and is too young to have had any supernovae \citep[i.e.\ age $<$4~Myr;][]{wang07}. Assuming a mass loss rate of $2\times 10^{-5}$~\Msol~yr$^{-1}$ and a terminal velocity of 2500~\kms\ for an O3-type star \citep{smith02}, this gives a wind power of $4\times 10^{37}$~erg~s$^{-1}$ per star or a total of $1.6\times 10^{38}$ for 4 stars. \textsc{Starburst}99 \citep{leitherer99} predicts a wind power of $10^{40}$~erg~s$^{-1}$ for a $10^6$~\Msun\ super star cluster, such as those typically found in starbursts. The total mechanical wind energy emitted from the Pismis 24 cluster is thus at least two orders of magnitude less than this, meaning that a lower level of turbulence in the surface layers of the pillar compared to those found in starburst environments is not unexpected.


\section{Conclusions} \label{sect:conc}
In this study we have examined in detail the gas conditions in the surface interaction layers of a gas pillar in NGC 6357 ionized by the star cluster Pismis 24 using high-resolution optical/near-IR IFU observations. Our observations show that the gas on the pillar's edge is denser, hotter, and more highly excited than its surroundings, providing clear evidence that the incoming radiation and winds from the cluster stars are strongly affecting the state of the ionized gas in these surface layers.

That we have been able to accurately decompose the H$\alpha$ line into multiple components has allowed us to examine the gas dynamics of this region in great detail. One of the primary aims of this study was to look for the presence of a broad emission component, and to investigate whether this component originates in turbulent mixing layers \citep[TMLs][]{slavin93, binette09} on the surface of the pillar set up by the shear flows of the winds from the young massive stars of the cluster Pismis 24, since this is the scenario suggesed by \citet[][see also \citealt{westm07b, westm07c, westm09a, westm09b}]{westm07a} to explain the broad emission components found in the starburst galaxies NGC 1569 and M82. On the pillar we find clear evidence of both a narrow ($\sim$20~\kms) and broad (50--150~\kms) component; away from the pillar we also see a much brighter, redshifted narrow component. The similarity of the broad C2 component velocities to the velocity of the narrow line component emitted at the location of the pillar leads us to conclude that the turbulent, C2-emitting gas must be associated with the pillar structure, whereas the redshifted narrow component originates in the background gas. That the velocities of the components associated with the pillar are consistent with those of the CO gas, and the fact that the broadest C2 line widths are found to follow the shape of the eastern pillar edge, lends further support to the connection between C2 and turbulent ionized gas within the pillar surface layers. We can therefore assert that the mechanical energy of the stellar winds impacting on the pillar surface is driving a turbulent mixing layer, and it is from this layer that the broad component is being emitted.

The pillar's edge is also clearly defined by dense ($\lesssim$5000~\cmt), hot ($\gtrsim$20\,000~K), and excited (via the [N\two]/H$\alpha$ and [S\two]/H$\alpha$ ratios) gas conditions, implying the presence of a D-type ionization front propagating into the pillar surface. Although there must be both photoevaporation outflows produced by the ionization front, and mass-loss through mechanical ablation (stripping), we see no evidence for any significant bulk gas motions on or around the pillar. We postulate that the evaporated/ablated gas must be rapidly heated before being entrained (i.e.\ the gas heating time is fast compared to outflow time).

Although the location of the broadest H$\alpha$ C2 emission follows the shape of the eastern pillar edge, it continues from the southern tip along a diagonal strip extending in the south-west direction. This extension is coincident with a faint protrusion to the tip of the pillar seen in the \textit{HST}/ACS F658N image. We conclude that it is either a part of the original molecular cloud that remains to be eroded or that it has been formed through a hydrodynamical instability in the outflowing material, and the geometry is such that it is shielding the western edge of the main pillar body from the oncoming winds therefore preventing the development of a TML on its surface.

Finally we have discussed the implications of our findings to what we have previously observed in starburst galaxies. The C2/C1 flux ratio is, in general, similar to what we have found in these energetic extragalactic environments, suggesting that the TMLs in both cases are of similar importance. That the C2 widths are narrower (50--150~\kms\ \textit{vs.} 150--350~\kms), however, could result from the fact that the Pismis 24 cluster contains only a few O stars and is too young to have had any supernovae, so the total mass flux and therefore mechanical wind energy is at least two orders of magnitude lower than that from a typical young massive star cluster such as those found in starburst galaxies. This effect results in a lower level of turbulence in the surface layers of the pillar.

To further investigate the relationships between the different gas components within the pillar, its surface and the surrounding gas more fully, it would be important to derive excitation, density and temperature diagnostics for each of the individual line components (C1, C2 and C3). This we have not been able to do since our observations of the faintest lines required for these diagnostics were not of a sufficient S/N. Thus, we intend to obtain deeper observations of this object in the future to build on the work presented here.

\section*{Acknowledgements}
We thank Mark Gieles for his help preparing and executing the observations. MSW would also like to thank Matt Redman and Mike Barlow for their careful reading of earlier versions of the manuscript and useful discussions regarding the nature of the observed wind-clump interactions. JSG's research was partially funded by the National Science Foundation through grant AST-0708967 to the University of Wisconsin-Madison.

\bibliographystyle{mn2e}
\bibliography{/Users/msw/Documents/work/references}

\bsp
\label{lastpage}
\end{document}